\documentclass[aps,pre,amsmath,amssymb,onecolumn,showkeys,showpacs,dvipdfmx]{revtex4-1}
\usepackage{amsmath}
\usepackage{amssymb}
\usepackage{bm}
\usepackage{dcolumn}
\usepackage[xdvi]{graphicx}
\usepackage{mathrsfs}
\usepackage{color}
\usepackage{mymacro}


\def\tr{{\rm Tr}}

\pacs{05.60.Gg, 05.60.-k, 03.65.Vf, 05.70.Ln, 44.10.+i}

\begin{document}
\title{Geometric fluctuation theorem for a spin-boson System}
\author{Kota L. Watanabe\footnote{Present address:
Works Application Inc., 1-12-32 Akasaka, Minato-ku, Tokyo 107-6019, Japan
} and Hisao Hayakawa \footnote{Corresponding author: e-mail: hisao@yukawa.kyoto-u.ac.jp}}
\affiliation{Yukawa Institute for Theoretical Physics, Kyoto University, Kitashirakawa-oiwake cho, Sakyo-ku, Kyoto 606-8502, Japan}
\begin{abstract}
 We derive an extended fluctuation theorem for geometric pumping of a spin-boson system under periodic control of environmental temperatures by using a Markovian quantum master equation.
We obtain the current distribution, the average current, and the fluctuation in terms of the Monte Carlo simulation.
To explain the results of our simulation we derive an extended fluctuation theorem.
This fluctuation theorem leads to the fluctuation dissipation relations but the absence of the conventional reciprocal relation.  
\end{abstract}

\maketitle

\section{Introduction}

Nonequilibrium processes are characterized by the existence of currents such as an electric current and a heat current from one reservoir
 to another reservoir. 
A nonequilibrium steady state under a steady current between reservoirs has been studied extensively.
In such a nonequilibrium steady system the current distribution $P_\tau(q_\tau)$ 
during time interval $\tau$ for the current $q_\tau$ from the left reservoir at temperature $\beta_L^{-1}$ and chemical potential $\mu_L$ 
to the right reservoir at $\beta_R^{-1}$ and $\mu_R$ 
satisfies the steady fluctuation theorem~ \cite{Evans}:
\begin{equation}
 \lim_{\tau \rightarrow \infty} \frac{1}{\tau} 
\ln \left\{\frac{P_{\tau}(q_{\tau})}{P_{\tau}(-q_{\tau})}\right\}
 = (\beta_L\mu_L -\beta_R \mu_R) \frac{q_{\tau}}{\tau} .
\end{equation}
This fluctuation theorem leads to various nonequilibrium relations among cumulants of the current such as the fluctuation dissipation relation (FDR) and the reciprocal relation.

There exists a different type current from the steady one, called the geometric pumping current in a mesoscopic system. 
The geometric pumping process is realized by the modulation of several control parameters such as chemical potentials, gate voltages, and tunneling barriers. 
It is remarkable that there exists a net average current without dc bias in the geometric pumping. 
This phenomenon has been observed in various processes such as quantized charge transports\cite{Thouless,Niu,Avron,Kouwenhoven,Pothier,Fuhrer,Kaestner,Chorley,Andreev,Makhlin,Aleiner}, spin pumpings\cite{Mucciolo,Governale,Cota,Splettstoesser1,Riwar,Deus,Watson,Nakajima} and qubit manipulations\cite{Brandes}. 
Such systems can be described by classical master equations\cite{Parrondo,Usami,Astumian1,Sinitsyn1,Sinitsyn2,Astumain2,Rahav,Ohkubo,Ren,Sagawa,Chernyak1,Chernyak2} or quantum master equations\cite{Cota,Splettstoesser1,Riwar,Nakajima,Brandes,Renzoni,Splettstoeser2,Reckemann,Hiltscher,Yuge}.
It is recognized that the mechanism of geometric pumping originated from the geometrical phase known as the Berry-Sinitsyn-Nemenman (BSN) phase~\cite{Sinitsyn1,Sinitsyn2,Ohkubo,Ren,Sagawa,Yuge} corresponding to the Berry phase in quantum mechanics \cite{Berry}. 
Our previous work \cite{watanabe} extended the adiabatic treatment to the nonadiabatic one for the calculation of the geometric current as the effect of the nonadiabatic BSN phase.

In previous studies of geometric pumping current, so far, the fluctuation theorem related to the BSN phase has not been systematically discussed, though there exists the current fluctuation theorem originated from the steady bias even for an open system within the Markovian approximation~\cite{Esposito,Jarzynski04,Saito07,Talkner,Noh,Gaspard07}.
In geometric pumping it has been suggested that the conventional fluctuation theorem does not exist \cite{Ren}. 
Interestingly, nonadiabatic effects play essential roles in the discussion whether the fluctuation theorem for the geometric pumping exists.

In this paper, we study geometric pumping effects within the framework of the quantum master equation under the Markovian approximation.
To clarify the argument, we only focus on the simplest spin-boson system in which there is a spin in the system attaching the left and the right reservoirs consisting of bosons~\cite{Ren,watanabe,Wang17}.
We also slowly control the temperatures in reservoirs sinusoidally characterized by the angular frequency $\Omega$. 
As a result, there is a geometric current from the reservoir to the system.
We perform the Monte Carlo simulation of this system and derive an extended fluctuation theorem for the geometrical pumping to explain the results of our simulation.
Our extended fluctuation theorem indicates that there exists FDR, while the conventional reciprocal relation is no longer valid in this system.

The organization of this paper is as follows.
In Sec.\ II, we introduce the spin-boson model and explain the methods of the generalized quantum master equation with the aid of full counting statistics (FCS) to analyze this system.
In Sec.\ III, we explain the method and results of the Monte Carlo simulation of our system. We also derive the fluctuation theorem from a phenomenological argument to be consistent with our numerical results.
Section IV consists of three parts. 
In Sec.\ IV A, we derive general expressions of the current distribution and the fluctuation theorem. 
In Sec.\ IV B, we illustrate the existence of the FDR but the absence of the reciprocal relation.
In Sec.\ IV C, we apply our formulation to a spin-boson system for the check of the validity of the derived fluctuation theorem.
Finally, we discuss and summarize our results, in Sec.\ V. 
In Appendix A, we describe the rate function introduced in Sec.\ IV A and summarize its properties.
In Appendix B, we summarize the used equations of physical variables for our analysis in a spin-boson system.
In Appendix C, we review the fluctuation theorem, the FDR, and the reciprocal relation for a steady current.

\section{Model and Method}

In this section, we explain the model and the method used for our analysis. 
We consider a nonequilibrium process under a periodic modulation of parameters in an open system consisting of a target system  in contact with a left reservoir (with the subscript $\rm L$) and a right reservoir (with the subscript $\rm R$).
The Hamiltonian of the total system consists of the system Hamiltonian $\hat H_{\rm S}$, 
\begin{equation}
\hat H_{\rm S}=\frac{\hbar \omega_0}{2}\hat \sigma_z,
\end{equation}
the Hamiltonian $\hat H_{\rm B}^{\nu}$ for the $\nu$th reservoir $(\nu = \rm L,R)$,  
\begin{equation}
\hat H_{\rm B}^\nu=\sum_{\bm{k}} \hbar \omega_{\bm{k},\nu}\hat b_{\bm{k},\nu}^\dagger \hat b_{\bm{k},\nu},
\end{equation}
and the interaction Hamiltonian $\hat H_{\rm SB}^{\nu}$ between the $\nu$th reservoir and the system
\begin{equation}
\hat H_{\rm SB}^\nu= \hbar \hat \sigma_x \sum_{\bm{k}} g_{\bm{k},\nu}(\hat b_{\bm{k},\nu}+\hat b_{\bm{k},\nu}^\dagger),
\end{equation} 
where $\hat \sigma_x$ and $\hat \sigma_z$ are Pauli matrices, $\hat b_{\bm{k},\nu}$ and $\hat b_{\bm{k},\nu}^\dagger$ are, respectively, bosonic annihilation and creation operators at the wave vector $\bm{k}$ for the environment $\nu$, and
$\hbar \omega_0$ and $\omega_{\bm{k},\nu}$ are the energy level and the angular frequency of the bosonic environment $\nu$, respectively. 
Here, $g_{\bm{k},\nu}$ is the coupling strength, which is characterized by the spectral density function $\mathscr{D}_\nu(\omega) \equiv 2\pi \sum_{\bm{k}} g_{\bm{k},\nu}^2 \delta (\omega -\omega_{\bm{k},\nu})$. 
For later analysis we use the dimensionless linewidth $\Gamma=\mathscr{D}_\nu(\omega)/\omega_0$ which is assumed to be independent of $\nu$ and $\omega$.
We also assume that the bosonic reservoirs are always in equilibrium, which is characterized by the equilibrium density matrix
$\hat \rho^{\rm eq,\nu}_{\rm B}(\beta_{\nu}(t))=e^{-\beta_{\nu}(t)\hat H_{\rm B}^{\nu}}/Z$
 at the inverse temperature $\beta_{\nu}(t)$ and time $t$, 
where $Z={\rm Tr}e^{-\beta_\nu(t)\hat H_{\rm B}^{\nu}}$ with the trace ${\rm Tr}$.
We modulate the temperatures of the reservoirs periodically with the angular frequency $\Omega$ by denoting  $\vec{\alpha}(t) = \{ \beta_{\rm L}(t), \beta_{\rm R}(t) \}$.
The positive (negative) angular velocity $\Omega$ corresponds to the counterclockwise (clockwise) rotation in the parameter space.
We note that we do not control parameters in the system Hamiltonian such as the gate voltage and the potential barrier in this paper.
It is straightforward to generalize formulation to cover such situations, but we restrict our interest to such a simple setup.  

Let us adopt the generalized quantum master equation with the aid of the full counting statistics (FCS) method. 
The FCS method enables us to obtain the probability distribution of the dimensionless heat transfer $q_\tau$ normalized by $\hbar\omega_0$ from a reservoir to the system during a time interval $\tau$.
 When we perform the projective measurement on a dimensionless energy $\hat Q$ at time zero and time $\tau$, the corresponding outcomes are denoted by $Q_0$ and $Q_\tau$, respectively. 
 The cumulant-generating function 
\begin{equation}
S_\tau(\chi) := \ln \int dq_{\tau} e^{i\chi q_{\tau}}P_{\tau}(q_{\tau})
\end{equation}
 is obtained by the probability distribution function $P_{\tau}(q_{\tau})$ of the (dimensionless) heat transfer $q_{\tau}=Q_{\tau}-Q_0$, where $\chi$ is the counting field. 
Note that the counting field is a dimensionless quantity, because its conjugate field $q_\tau$ is the dimensionless heat transfer. 
 In this method, the cumulant-generating function is given by $S_{\tau}(\chi)=\tr \hat \rho^{\chi}(\tau)$, where $\hat \rho^{\chi}(\tau)$ is the generalized density matrix for the target system, which satisfies the generalized quantum master equation

\begin{eqnarray}
\df{}{t} \hat \rho^{\chi}(t) = \omega_0 \mathcal K^{\chi}_{\vec{\alpha}(t)}\hat \rho^{\chi}(t), \label{qme0}
\end{eqnarray}
where the superoperator $\mathcal K^{\chi}_{\vec{\alpha}(t)}$ has the periodicity from $\vec{\alpha}(t+\tau_p) = \vec{\alpha}(t)$ with the period of the modulation $\tau_p = 2\pi/|\Omega|$. 
Equation  (\ref{qme0}) is explicitly written as the Born-Markov quantum master equation
\begin{eqnarray}
\frac{d}{dt} \hat \rho^{\chi}(t)=-\frac{i}{\hbar}[\hat H_{\rm S}, \hat \rho^{\chi}(t)]  -\sum_{\nu=\rm L,R}\frac{1}{\hbar^2}\int_0^{\infty} d\tau {\rm Tr}_E [ \hat H_{ \rm SE}^\nu, [ \hat H_{\rm SE}^\nu(-\tau), \hat \rho_{\rm B}^{\rm eq,\nu}(\vec{ \alpha}(t)) \hat \rho^{\chi}(t)]_\chi ]_\chi, \label{bmqm}
\end{eqnarray}
where we have introduced $[\hat H,\hat A]_{\chi} := \hat H_\chi \hat A- \hat A \hat H_{-\chi}$ for an arbitrary operator $\hat A$, $ \hat H_\chi:= e^{i\chi \hat Q/2}\hat H e^{-i\chi \hat Q/2}$, $\rho_{\rm B}^{\rm eq,\nu}(\vec{ \alpha}(t)) := \hat \rho^{\rm eq,L}_{\rm B}(\beta_{\rm L}(t))\hat \rho^{\rm eq,R}_{\rm B}(\beta_{\rm R}(t)) $, and $\hat H_{\rm SE}^{\nu}(\tau) = e^{i(\hat H_{\rm S} +\hat H_{\rm E})\tau/\hbar}\hat H_{\rm SE}^{\nu} e^{-i(\hat H_{\rm S} +\hat H_{\rm E})\tau/\hbar}$. 
Thus, the explicit definition of the superoperator ${\cal K}_{\vec{\alpha}(t)}^\chi$ in Eq.~\eqref{qme0} is given by the right hand side of Eq.~\eqref{bmqm}.
Usually Eq.\ (\ref{bmqm}) is only valid for steady nonequilibrium situations, but Ref.\cite{watanabe} shows that Eq. (\ref{bmqm}) can be used for systems under slowly modulated parameters $\vec{\alpha}(t)$.
Note that Eq.\ (\ref{bmqm}) is applicable when the coupling constant, {\it \rm i.e.}
 the dimensionless bandwidth $\Gamma$, 
 is weak.  
We also note that $\hat \rho^{\chi}(\tau)$ is reduced to the usual density matrix at $\chi=0$ and satisfies $\hat \rho^{\chi}(\tau)^{\dagger} = \hat \rho^{-\chi}(\tau)$.

In the spin-boson system, the matrix representation of Eq.\ (\ref{bmqm}) is written as 
\begin{eqnarray}
\df{}{t}\kket{\rho^{\chi}(t)} = \omega_0 K^{\chi}(t) \kket{\rho^{\chi}(t)}, \label{QME3}
\end{eqnarray}
where $\kket{\rho^{\chi}(t)}$ is the vector satisfying $\kket{\rho^{\chi}(t)} = {}^T( \bra{0} \hat \rho^{\chi}(t)\ket{0}, \bra{0} \hat \rho^{\chi}(t)\ket{1}, \bra{1} \hat \rho^{\chi}(t)\ket{0}, \bra{1} \hat \rho^{\chi}(t)\ket{1})$ with the notation of the transverse(superscript $T$) corresponding to the up spin state $|1\rangle$ and the down spin state $|0\rangle$, 
and the matrix $K^{\chi}(t)$ is given by
\begin{equation}
K^{\chi}(t) = 
\begin{pmatrix}
 a_1(t) & 0 & 0 & a_2^{\chi}(t) \\
0 & c_1^{\chi} (t) & c_2^{\chi}(t) & 0 \\
0 & c_2^{\chi} (t)^* & c_1(t)^* & 0 \\
a_3^{\chi}(t)& 0 & 0 & a_4(t)   \\
 \end{pmatrix}
.\label{Markov-K}
\end{equation}
Here, the explicit forms of the matrix elements are given by
\begin{eqnarray}
a_1(t) &=&-\frac{1}{\omega_0}\sum_{\nu=\rm L,R} \int_0^{\infty}d\tau \{ \Phi_{1,\nu}(\vec{\alpha}(t),\tau)e^{-i\omega_0\tau} + \Phi_{1,\nu}^{*}(\vec{\alpha}(t),\tau)e^{i\omega_0\tau}\} = -\Gamma[ n_{\rm L}(t) + n_{\rm R}(t)], \label{a1}
 \\
a_2^{\chi}(t) &=& \frac{1}{\omega_0}
\sum_{\nu=\rm L,R} \int_0^{\infty}d\tau\{ \Phi_{2,\nu}^{\chi}(\vec{\alpha}(t),\tau)e^{-i\omega_0\tau} + \Phi_{3,\nu}^{\chi}(\vec{\alpha}(t),\tau)e^{i\omega_0\tau}\} = \Gamma[ 1+ n_{\rm L}(t)+ (1+ n_{\rm R}(t))e^{i\chi \hbar \omega_0}],\label{a2} 
\\
a_3^{\chi}(t) &=& \frac{1}{\omega_0}
\sum_{\nu=\rm L,R} \int_0^{\infty}d\tau\{ \Phi_{2,\nu}^{\chi}(\vec{\alpha}(t)\tau)e^{i\omega_0\tau} + \Phi_{3,\nu}^{\chi}(\vec{\alpha}(t),\tau)e^{-i\omega_0\tau}\} = \Gamma[ n_{\rm L}( t)+n_{\rm R}(t)e^{-i\chi\hbar \omega_0}],\label{a3} 
\\
a_4(t) &=& -\frac{1}{\omega_0}
\sum_{\nu=\rm L,R}\int_0^{\infty}d\tau \{ \Phi_{1,\nu}(\vec{\alpha}(t),\tau)e^{i\omega_0\tau} + \Phi_{1,\nu}^{*}(\vec{\alpha}(t),\tau)e^{-i\omega_0\tau}\} = - \Gamma[ 2+ n_{\rm L}(t) + n_{\rm R}(t) ], \label{a4} 
\\
c_1(t) &=& i -\frac{1}{\omega_0}\sum_{\nu=\rm L,R} \int_0^{\infty}d\tau \{ \Phi_{1,\nu}(\vec{\alpha}(t),\tau) + \Phi_{1,\nu}^{*}(\vec{\alpha}(t),\tau)\}e^{-i\omega_0\tau}, 
\\
c_2^{\chi} (t) &=& \frac{1}{\omega_0}
\sum_{\nu=\rm L,R} \int_0^{\infty}d\tau\{ \Phi_{2,\nu}^{\chi}(\vec{\alpha}(t),\tau) + \Phi_{3,\nu}^{\chi}(\vec{\alpha}(t),\tau)\}e^{i\omega_0\tau} ,
\end{eqnarray}
where correlation factors $\Phi_{1,\nu}, \Phi_{2,\nu}$, and $\Phi_{3,\nu}$ are given as
\begin{eqnarray}
\Phi_{1,\nu}(\vec{\alpha}(t),\tau)&=&\sum_{\bm{k}} g_{\bm{k},\nu}^2 \{ \expec{\hat b_{\bm{k},\nu}^{\dagger}\hat b_{\bm{k},\nu}}_{\beta_{\nu}(t)}e^{i\omega_{\bm{k},\nu}\tau} + \expec{\hat b_{\bm{k},\nu}\hat b_{\bm{k},\nu}^{\dagger}}_{\beta_{\nu}(t)}e^{-i\omega_{\bm{k},\nu}\tau}\}, \\
\Phi_{2,\nu}^{\chi}(\vec{\alpha}(t),\tau)&=
&\sum_{\bm{k}} g_{\bm{k},\nu}^2
\{ \expec{\hat b_{\bm{k},\nu}^{\dagger}\hat b_{\bm{k},\nu}}_{\beta_{\nu}(t)}
e^{-i\omega_{\bm{k},\nu}\tau - i\hbar\omega_{\bm{k},\nu}\chi_{\nu}}
 + \expec{\hat b_{\bm{k},\nu}\hat b_{\bm{k},\nu}^{\dagger}}_{\beta_{\nu}(t)}e^{i\omega_{\bm{k},\nu}\tau + i\hbar\omega_{\bm{k},\nu}\chi_{\nu}}\}, \\
\Phi_{3,\nu}^{\chi}(\vec{\alpha}(t),\tau) &=&\sum_{\bm{k}} g_{\bm{k},\nu}^2\{\expec{\hat b_{\bm{k},\nu}^{\dagger}\hat b_{\bm{k},\nu}}_{\beta_{\nu}(t)}e^{i\omega_{\bm{k},\nu}\tau - i\hbar\omega_{\bm{k},\nu}\chi_{\nu}} + \expec{\hat b_{\bm{k},\nu}\hat b_{\bm{k},\nu}^{\dagger}}_{\beta_{\nu}(t)}e^{-i\omega_{\bm{k},\nu}\tau + i\hbar\omega_{\bm{k},\nu}\chi_{\nu}}\}
\end{eqnarray}
with the aid of the canonical ensemble $ \expec{ \cdots }_{\beta_{\nu}} := \tr_B\{ \cdots \hat \rho^{\rm eq,\nu}_{\rm B}(\beta_{\nu})\} $ and the Bose-Einstein distribution $n_{\nu}(t) = (e^{\beta_{\nu}(t) \hbar \omega_0} -1 )^{-1}$.
As can be seen from Eq.\ (\ref{Markov-K}), the diagonal part is independent of the off-diagonal part. 
Then, we consider the time evolution only of the diagonal part. 
We also assume that the time evolution of $\vec{\alpha}(t)$ satisfies
\begin{eqnarray}\label{T(t)}
\beta_{\rm L}(t)^{-1}&=&  T_{0}+ T_A \cos(\Omega t+\pi/4) , \nonumber\\
\beta_{\rm R}(t)^{-1}&=&  T_{0}+ T_A \sin(\Omega t+\pi/4) ,
\end{eqnarray}
where $T_{0}$ and $T_A$ are, respectively, the average temperature and the amplitude of the modulation

\section{Numerical simulation}

In this section, we show the results of our Monte Carlo simulation for the cumulants associated with the heat transfer.
As a result, we obtain the current distribution and the fluctuation theorem.

\subsection{How to perform the Monte Carlo simulation}

In this section, we explain the method to perform our simulation.
We begin with the Fourier transform of Eq.\ (\ref{qme0}) :
\begin{eqnarray}
\df{}{t}\tilde \rho^{q}(t) = \omega_0 \int dq' \tilde{\mathcal K}^{q-q'}_{\vec{\alpha}(t)}\tilde \rho^{q}(t), \label{qqme}
\end{eqnarray}
where the Fourier transform $\tilde A^q $ of $A^{\chi}$ is defined by
\begin{eqnarray}
\tilde A^q = \frac{1}{2\pi}\int_{-\infty}^{\infty} d\chi e^{-i\chi q} A^{\chi}.
\end{eqnarray}
Note that the label $q$ corresponds to the heat transfer. Because $\tilde \rho^{q}(t)$ is Hermitian, we can use the spectral decomposition 
\begin{eqnarray}
\tilde \rho^{q}(t) = \sum_{m_t}p^{q}_{m_t}(t)\ket{m_t}\bra{m_t}. \label{spd} ,
\end{eqnarray}
where $\ket{m}$ takes 
$\ket{1}$ or  
$\ket{0}$.
Because the spectrum coefficient is bounded as $0 \leq p^{q}_{m_t} \leq 1$,  $p^{q}_{m_t}$ can be regarded as the probability for the state $(m,q)$. 
With the aid of Eq.\ (\ref{spd}), the quantum master equation [Eq.\ (\ref{qqme})] is reduced to a classical master equation as
\begin{eqnarray}\label{eq22}
\dot p^{q}_m(t) = \sum_{m'}\int dq' W_t(m,q|m',q')p^{q'}_{m'}(t), \label{2mas} 
\end{eqnarray}
where $\dot{p}_m^q(t)$ expresses $d p^q_m(t)/d\tilde{t}$ with $\tilde t = \omega_0 t$, and $W_t(m,q|m',q')$ is the transition rate from $(m',q')$ to $(m,q)$ defined by
\begin{eqnarray}
W_t(m,q|m',q') = \tr\{ \ket{m}\bra{m}\hat {\mathcal K}^{q-q'}_{\vec{\alpha}(t)}\ket{m'}\bra{m'}\},
\end{eqnarray}
and satisfies the preservation of probability 
\begin{eqnarray}
\sum_{m,q} W_t(m,q|m',q') = 0. 
\label{pp}
\end{eqnarray}


\subsection{Application to the spin-boson system}

Now, let us apply the above method to the spin-boson system.
From Eq.\ (\ref{Markov-K}), the matrix of $W_t(m,q|m',q')$ is rewritten as 
\begin{eqnarray}
& & \ \ \ m = 0 \ \ \ m = 1 \nonumber \\
W_t(m,q|m',q') &=& \begin{pmatrix}
\tilde a_1^{q-q'}(t) & \tilde a_2^{q-q'}(t)  \\
\tilde a_3^{q-q'}(t) & \tilde a_4^{q-q'}(t)
\end{pmatrix} \begin{matrix}m' = 0 \\m' = 1 \end{matrix}\ ,
\end{eqnarray}
where the matrix elements are explicitly given by
\begin{eqnarray}
\tilde a_1^{q}(t) &=& -\Gamma[n_{\rm L}(t)+n_{\rm R}(t)]\delta(q),  \\
\tilde a_2^q(t) &=& \Gamma[(1+n_{\rm L}(t))\delta(q) + (1+n_{\rm R}(t))\delta(q-\hbar \omega_0)], \\
\tilde a_3^q(t) &=& \Gamma[ n_{\rm L}(t)\delta(q) +  n_{\rm R}(t)\delta(q+\hbar \omega_0)], \\
\tilde a_4^q(t) &=&- \Gamma[ 2+n_{\rm L}(t)+n_{\rm R}(t) ]\delta(q).
\end{eqnarray}
Introducing $p_0^{q} = \bra{0}\tilde \rho^{q}\ket{0}$ and $p_1^{q}= \bra{1}\tilde \rho^{q}\ket{1}$, master equation \eqref{eq22} is rewritten as
\begin{eqnarray}
\dot p^q_0(t) &=& \Gamma n_{\rm L}(t)[p_1^q(t)-p_0^q(t)] +\Gamma n_{\rm R}(t)[p_1^{q-1}(t)-p_0^q(t)], \\
\dot p^q_1(t) &=& \Gamma(1+ n_{\rm L}(t))[p_0^q(t)-p_1^q(t)] +\Gamma (1+n_{\rm R}(t))[p_0^{q+1}(t)-p_1^q(t)] .
\end{eqnarray}
We note that $q\pm 1$ corresponds to the heat increment $\pm \hbar\omega_0$ from the original state in the physical unit.
With the aid of the Poisson noise, the time evolutions of stochastic variables $(\hat m(t), \hat q(t) ) $ are, respectively, given by
\begin{eqnarray}
\hat m(t+\Delta t) &=& \hat m(t) + \delta_{\hat m(t),0}[\hat \xi_{ \Gamma n_{\rm L}(t)}^P(t)+\hat \xi_{ \Gamma n_{\rm R}(t)}^P(t)] \Delta t \nonumber \\
& & + \delta_{\hat m(t),1}[\hat \xi_{ \Gamma (1+n_{\rm R}(t))}^P(t)+\hat \xi_{ \Gamma (1+n_{\rm L}(t))}^P(t)]\Delta t \ \ \ {\rm mod} \ 2  \label{mte}\\
\hat q(t+\Delta t) &=& \hat q(t) + 
[\delta_{\hat m(t),1}\hat \xi_{ \Gamma n_{\rm R}(t)}^P(t)-\delta_{\hat m(t),0}\hat \xi_{ \Gamma (1+n_{\rm R}(t))}^P(t)]\Delta t, \label{qte}
\end{eqnarray}
where $\hat \xi_{ \lambda}^P(t)$ is the Poisson noise with the intensity $y^* =1$. Then, we obtain the probability distribution $P(\hat q(\tau_p) = q)$. 
When we consider the time-reversal path, we replace $\Delta t$ by $ -\Delta t$. 

\subsection{The results of our simulation}

\begin{figure}[htbp]
\centering
\includegraphics[scale=0.7]{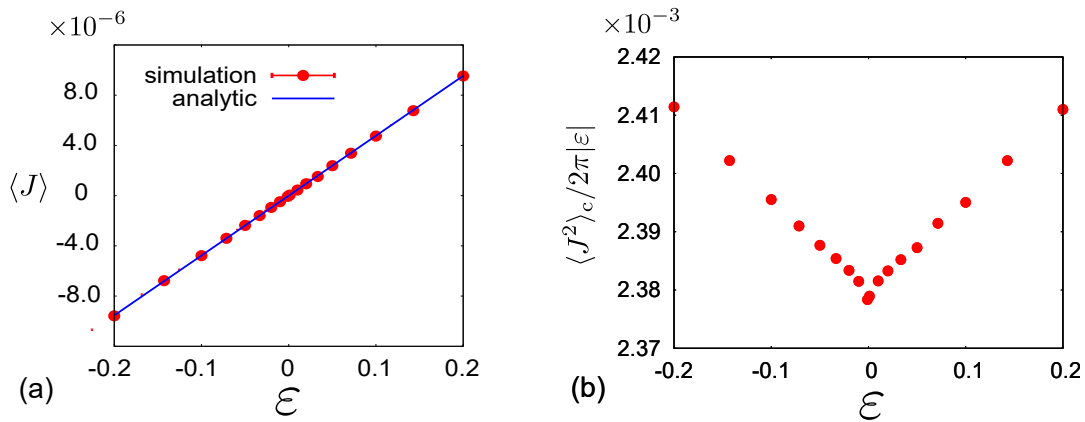}
 \caption{(Color online)
  The plots of (a) the  average $\expec{J}$ and 
  (b) the fluctuation $ \expec{J^2}_c/(2\pi|\varepsilon|)$
   obtained from the Monte Carlo simulation of Eqs.\ (\ref{mte}) and (\ref{qte}),
    where the number of samples is 20,000,000 for each $\varepsilon$.
The parameters are modulated by $(\beta_{\rm L}(t)\hbar \omega_0)^{-1} = 0.7 + 0.35\cos{(\Omega t +\pi/4)}$, $(\beta_{\rm R}(t)\hbar \omega_0)^{-1} = 0.7 + 0.35\sin{(\Omega t +\pi/4)}$, and $\Gamma = 0.01$. We have eliminated the effects of  initial conditions by omitting the data in the transient processes to reach the steady states. 
The best fitting of the behavior for the fluctuation is 
$ \expec{J^2}_c \propto |\varepsilon|^{0.85}$. 
}
 \label{fig1}
\end{figure}

In this subsection, we present the results of our numerical simulation for quantities associated with the dimensionless current $J$ defined by $J: =  q_{\tau_p}/(\tau_p \omega_0 \Gamma)=\varepsilon \tilde{q}$, 
where $\varepsilon:=\Omega/(\omega_0\Gamma)$ and $\tilde{q}:=|q_{\tau_p}|/(2\pi)$. 
We set $\Gamma = 0.01$ in this paper. 
Our numerical simulation is performed for sufficiently small $|\varepsilon|$ as shown in Figs.\ 1$-$4.
Figures 1 (a) and (b) are plots of the average current $\expec{J}$ and the fluctuation $\expec{J^2}_{c}=\expec{J^2} - \expec{J}^2$ against $\varepsilon$. 
In Fig.1(a), the solid line is the analytic expression obtained in past studies\cite{Ren,watanabe}. 
This good agreement between the theory and the simulation ensures the validity of our Monte Carlo simulation.
From Fig. 1(b), the behavior of $\expec{(q_{\tau_p}-\expec{q_{\tau_p}})^2}/(\tau_p \omega_0 \Gamma)=\expec{J^2}_c/(2\pi|\varepsilon|)$ seems to be proportional to $|\varepsilon|$, though the origin of the singularity in the differentiation at $\varepsilon=0$ is not identified.
Figure 2 confirms that the current distribution approximately satisfies the Gaussian distribution. 
Nevertheless, there is a systematic deviation from the Gaussian as shown in Fig.~\ref{fig4}, where
the current distribution is given by 
\begin{eqnarray}
P_{\varepsilon}(J) \sim \exp{\left[ -|\varepsilon|^{-1} \left\{ \frac{a(|\varepsilon|)}{2} (J-\expec{J} )^2 + \frac{b(|\varepsilon|)}{4} (J-\expec{J} )^4  +  \cdots \right\} \right]}. 
\label{gaussdev}
\end{eqnarray}
We estimate $b(0) \simeq 5.4 \times 10^6$ 
from Fig.~\ref{fig4}.

\begin{figure}[htbp]
\centering
\includegraphics[scale=0.7]{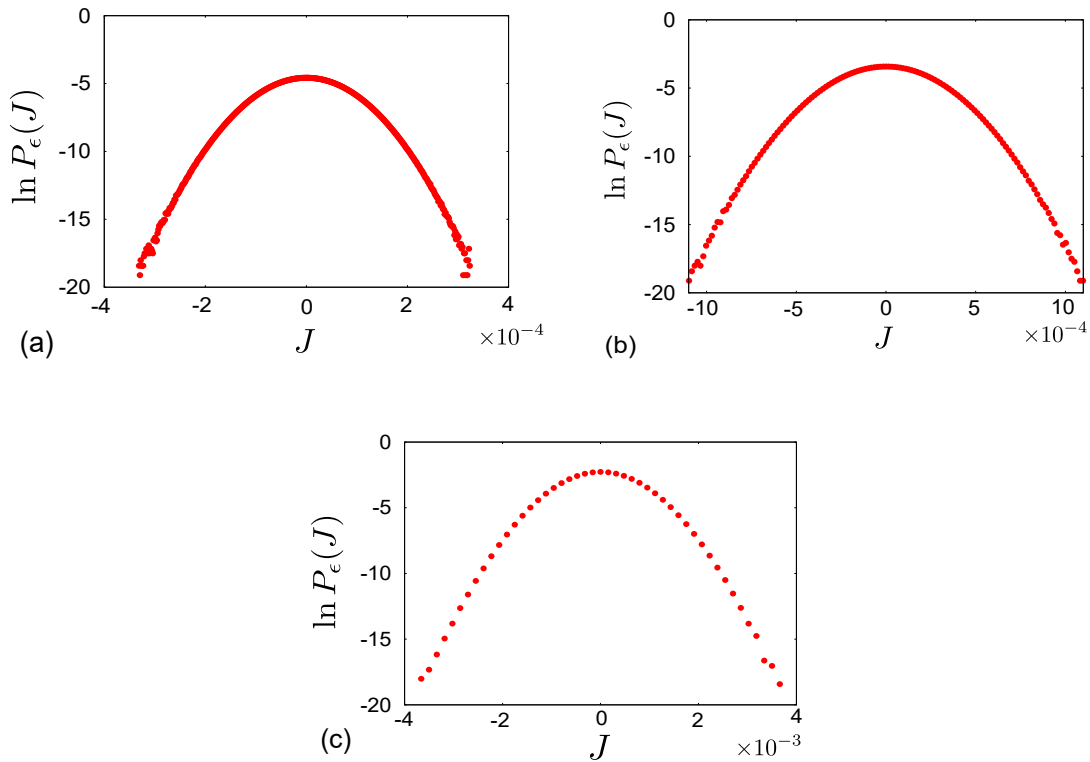}
 \caption{
 (Color online) 
 The plots of the current distributions for (a) 
 $\varepsilon=0.001$, (b) $\varepsilon=0.01$, and (c) $\varepsilon=0.1$, 
 where the set of the other parameters is equivalent to that used in Fig. \ref{fig1}.
  The average $\expec{J}$ is small at $O(|\varepsilon|)$ but it is the finite value from 
  Fig. 1(a) and Eq. (\ref{<J>}).}
 \label{fig2}
\end{figure}
\begin{figure}[htbp]
\centering
  \includegraphics[scale =0.5]{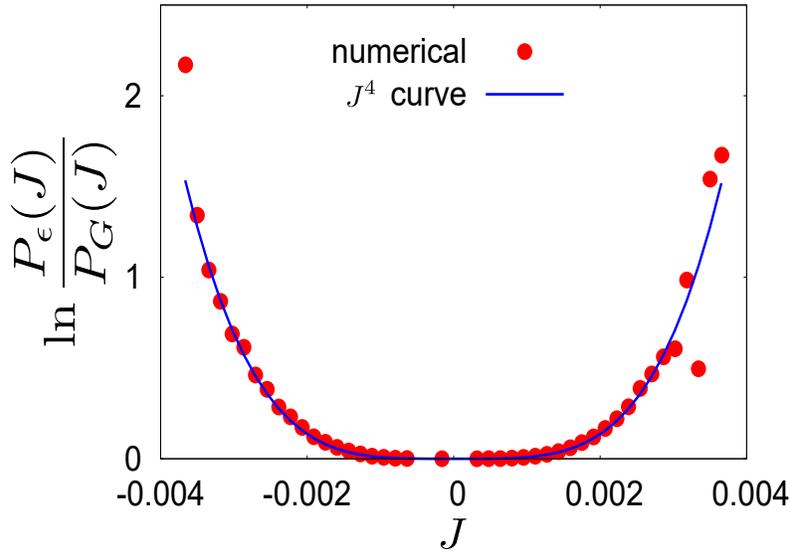}
 \caption{
 (Color online) 
 The plot of the deviation from the Gaussian $P_G(J)
 \sim \exp[-a(0)(J-\langle J\rangle)^2/(2|\varepsilon|) ]
$
  for $\Omega/\Gamma = 0.1$. 
  The line represents the fourth-order fit curve. 
  From this figure we can estimate the coefficients of
   $J^4$ in Eq.\ \eqref{gaussdev} as $b(0) \simeq 5.4 \times 10^6$
. 
}
 \label{fig4}
\end{figure}

Together with the nonadiabatic current reported in Ref.~\cite{watanabe},
\begin{equation}\label{<J>}
\langle J \rangle =\alpha \varepsilon +\beta \varepsilon^3+\cdots ,
\end{equation}
we can rewrite Eq.~\eqref{gaussdev} as
\begin{eqnarray}
\lim_{| \varepsilon| \rightarrow 0 } |\varepsilon| \ln \frac{P_{\varepsilon}(J)}{P_{\varepsilon}(-J)}  = \varepsilon \{A_1 J + A_3 J^3  + \cdots\} + \varepsilon |\varepsilon| B J + \cdots  + O(\varepsilon^3), \label{ft1}
\end{eqnarray}
where the expansion coefficients satisfy the relations
\begin{eqnarray}
A_1 &=& 2 \alpha  a(0) , \label{coa} \\
A_3 &=& 2 \alpha  b(0) , \label{cob}\\
B &=& 2 \alpha a'(0), \label{coex}
\end{eqnarray}
We note that $A_1$, $A_3$, \dots, and $B$ are independent of $\varepsilon$.  
It should be noted that $A_1$ is not a small quantity in our situation, while the affinity $\mathcal A$ for the steady fluctuation theorem in Appendix C is small. 

\begin{figure}[htbp]
\centering
\includegraphics[scale=0.7]{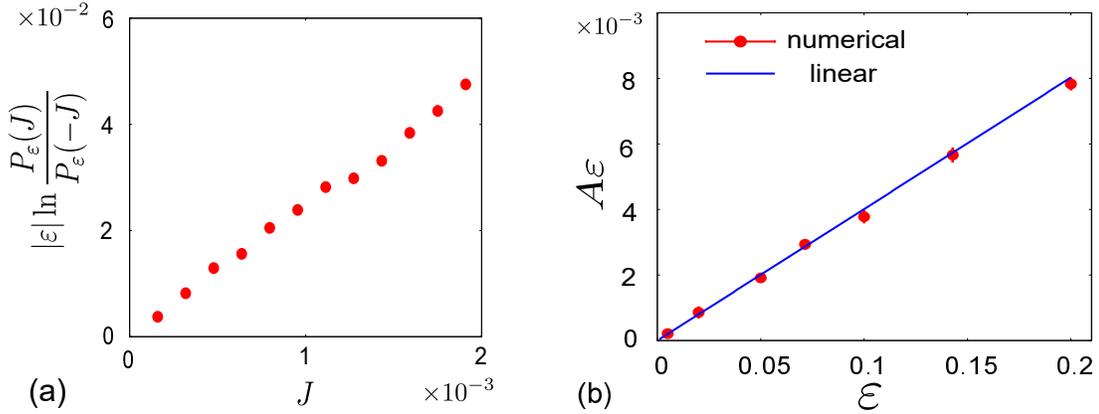}
 \caption{(Color online) (a) 
 The plot of Eq.\ (\ref{ft1}) versus $J$ in $\varepsilon = 0.1$ corresponding to Fig. 2(c). The set of the other parameters is equivalent to that used in Fig. \ref{fig1}. Dots in (b) are calculated by the arithmetic average of $|\varepsilon| \ln{[P_{\varepsilon}(J)/P_{\varepsilon}(-J)]}/J$ defined as $\sum_{|J| \leq J_{\max}, J\neq 0} \left\{ |\varepsilon| \ln{[P_{\varepsilon}(J)/P_{\varepsilon}(-J)]}/J \right\} /N$, where $N$ is the number of $J \in [-J_{\max},J_{\max}]$ and $J_{\max}$ is the threshold excluding large deviations
 of $P_{\varepsilon}(\pm J)$, which are $J_{\max} = 3.9 \times 10^{-4}, N = 106$ for $\varepsilon = 0.005$,  $J_{\max} = 1.3 \times 10^{-3}, N = 82$ for $\varepsilon = 0.02$, $J_{\max} = 2.1 \times 10^{-3},N=52$ for $\varepsilon = 0.05$, $J_{\max} = 2.3 \times 10^{-3},N=40$ for $\varepsilon = 0.07$, $J_{\max} = 2.9 \times 10^{-3},N=36$ for $\varepsilon = 0.1$, $J_{\max} = 3.7 \times 10^{-3}, N = 32$ for $\varepsilon = 0.14$, and $J_{\max} = 3.9 \times 10^{-3}, N =24$ for $\varepsilon = 0.2$.
From this plot, we obtain $A \simeq 4 \times 10^{-2}$. }
 \label{fig3}
\end{figure}

Equation~\eqref{ft1} is the fluctuation theorem for the geometric current. 
Indeed,  Figs.~\ref{fig4} and \ref{fig3}  are consistent with Eq.~\eqref{ft1}, though non-Gaussian contribution proportional to $J^3$ is not clearly visible in Fig.~\ref{fig3}. 
Note that Fig.~\ref{fig3} (b) is plotted by the arithmetic average for $J$, as explained in the caption of Fig.~\ref{fig3}. 
The solid line in Fig. \ref{fig3} (b) is the analytic expression by a linear fitting function. 
From this figure, $A_1$ can be estimated as $A_1 \simeq 4 \times 10^{-2}$.

\section{ Theory of Geometric current distribution}

Now, let us give the theoretical basis of the extended fluctuation theorem, Eq.~\eqref{ft1}, for the geometric pumping process which is phenomenologically derived. 
In Sec. IV A, we express
the formulas in relation to the fluctuation theorem in the spin-boson system under periodic modulation of the environmental temperatures. 
In Sec. IV B, we derive relations among the cumulants of the currents. 
Here, we confirm that the conventional FDR is held, while we demonstrate the absence of the conventional reciprocal relation.
In Sec. IV C, we compare our theoretical results with the numerical results presented in the previous section.

\subsection{General formulation}

In this section, let us derive formal expressions of the current distribution and the fluctuation theorem for the geometric pumping process.
We introduce $\lambda_0^{\chi}(t)$ as the largest eigenvalue with the superoperator $\mathcal K^{\chi}_{\vec{\alpha}(t)}$ in Eq.\ (\ref{qme0}).
We also introduce $\bbra{l_0^{\chi}(t)}$ and $\kket{r_0^{\chi}(t)}$ for the left and right eigenvectors of $\mathcal K^{\chi}_{\vec{\alpha}(t)}$, respectively, corresponding to $\lambda_0^{\chi}(t)$. 
The largest eigenvalue $\lambda_0^{\chi}(t)$ satisfies Levitov-Lesovik-Gallavotti-Cohen (LLGC) symmetry \cite{Levitov,Gallavotti,Yuge2013}:
\begin{eqnarray}
\lambda_0^{\chi}(t) = \lambda_0^{-\chi+i\Delta \alpha(t)}(t), \label{sym}
\end{eqnarray}
where $\Delta \alpha(t)$ denotes the bias parameter 
$\Delta\alpha(t)= \hbar\omega_0(\beta_{\rm L}(t) -\beta_{\rm R}(t))$.
Note that the symmetry relation in Eq.\ (\ref{bmqm}) was proved in Ref. \cite{Yuge2013}.
(See also Appendix B for the proof of symmetry relation (\ref{sym}) in the spin-boson system.)
If parameters are independent of time, Eq.\ (\ref{sym}) directly leads to the steady fluctuation theorem. 
When $\Delta \alpha(t)$ depends on $t$, the conventional fluctuation theorem is no longer valid.
From Eq.\ (\ref{sym}), $\lambda_0^{\chi}(t)$ should be an even function of $\chi-i\Delta \alpha(t)/2$.

Under the periodic modulation of parameters, we assume that the cumulant-generating function consists of
\begin{eqnarray}
S_{\tau_p}(\chi) = S_{\tau_p}^{\rm dyn}(\chi) + S_{\tau_p}^{\rm geo}(\chi), \label{cumu}
\end{eqnarray}
where $S_{\tau_p}^{\rm dyn}(\chi)$ is the contribution of the dynamical phase and $S_{\tau}^{\rm geo}(\chi)$ is the contribution of the geometric phase (Ref.\ \cite{Ren}) including nonadiabatic contributions.
The cumulant-generating function of the dynamical phase is given by
\begin{eqnarray}
S_{\tau_p}^{\rm dyn}(\chi)|\varepsilon| :=  \overline{\lambda_0^{\chi}}
:=
\frac{1}{2\pi}\int_{\theta_0}^{\theta_0 + 2\pi} d\theta \lambda_0^{\chi}(\theta) , \label{dynp}
\end{eqnarray}
where $\theta$ and $\theta_0$ are the phase of the modulation and the initial phase to remove the initial condition dependence, respectively. 
For later convenience, we have introduced $\overline{\mathcal A}:=\frac{1}{2\pi}\int_{\theta_0}^{\theta_0+2\pi}d\theta {\mathcal A}(\theta)$.
On the other hand, to be consistent with the numerical results in the previous section, we assume that the geometrical cumulant-generating function is given by
\begin{eqnarray}
S_{\tau_p}^{\rm geo}(\chi)|\varepsilon| 
=\varepsilon \overline{v_{\rm adi}^{\chi}} + |\varepsilon| \overline{v_{\rm ex}^{\chi}}+ \sum_{ j = 2}^{\infty} \varepsilon^j \overline {v_{{\rm na},j}^{\chi}}. \label{geoexp}
\end{eqnarray}
Here, $\overline{v_{\rm adi}^{\chi}}$ in Eq.\ (\ref{geoexp}) is the BSN phase represented by
\begin{eqnarray}
\overline{v_{\rm adi}^{\chi}} &=&
 - \frac{1}{2\pi}\int_{\theta_0}^{2\pi+\theta_0} d\theta \bbra{l_0^{\chi}(\theta)}\df{}{\theta}\kket{r_0^{\chi}(\theta)}.
\end{eqnarray}
It is a key assumption to introduce $\overline{v_{\rm ex}^{\chi}}$ in Eq.\ (\ref{geoexp}) to be consistent with Fig.1(b).
We also assume that the nonlinear contributions of $\varepsilon$ can be expanded as a series of $\varepsilon^j$ with $j\ge 2$. 
Thus, $\overline{v_{{\rm na}, j}^{\chi}}$ is the expansion coefficient of $\varepsilon^j$.
Note that the cumulant-generating function must be invariant under the transformation $\Omega\to -\Omega$ and $\chi \to - \chi$, because the physical process should be independent of the sign of $\Omega$ 
if the basic equation \eqref{bmqm} is invariant under this transformation. 
Then, the parities of the geometric factors are summarized as 
\begin{eqnarray}
\overline{v_{\rm adi}^{\chi}} &=& -\overline{v_{\rm adi}^{-\chi}}, \label{adpar}\\
\overline{v_{\rm ex}^{\chi}} &=& \overline{v_{\rm ex}^{-\chi}}, \label{expar}\\
\overline{v_{{\rm na},j}^{\chi} }&=& (-1)^j \overline{v_{{\rm na},j}^{-\chi}}. \label{napar}
\end{eqnarray}

We now derive the fluctuation theorem under the periodic modulation of parameters under the condition $\overline{\Delta \alpha(t)} = 0$. 
Note that the dynamical phase plays dominant roles if $\overline{\Delta\alpha(t)}\ne 0$.
Let us begin with the formal expression for the probability distribution of a pumping current 
\begin{eqnarray}
P_{\varepsilon}(J) =|\varepsilon|^{-1}\int_{-\infty}^{\infty} \frac{d\chi}{2\pi}e^{-|\varepsilon|^{-1} [i\chi J - |\varepsilon|S_{\tau_p}(\chi)] } . \label{prob}
\end{eqnarray}
We evaluate this integration for large $|\varepsilon|^{-1}$. 
For this purpose, we introduce the rate function $I(J)$ defined as
\begin{eqnarray}
I(J) := \max_{\chi}[i\chi J -\overline{\lambda_0^{\chi}} ] =i\chi_c(J) J -  \overline{\lambda_0^{\chi_c(J)}} ,
\end{eqnarray}
where ${\rm max}_\chi$ expresses the operation to select the maximum value within possible $\chi$.
We also introduce the residual term $g(\chi)$ defined as
\begin{eqnarray}
g(\chi) = i\{\chi-\chi_c(J) \} J -2\pi \{\overline{\lambda_0^{\chi}}-\overline{\lambda_0^{\chi_c(J)}} \}, \label{defg}
\end{eqnarray}
where $\chi_c(J)$ expresses the saddle point, i.e., 
the solution of the equation $J= \partial_{i\chi}\overline{\lambda_0^{\chi}} |_{\chi=\chi_c}$. The rate function $I(J)$ is an even function of $J$ satisfying $I(J) = I(-J)$ (see Appendix A).

Then, Eq.\ (\ref{prob}) can be rewritten as
\begin{eqnarray}
P_{\varepsilon}(J)|\varepsilon|  = e^{-|\varepsilon|^{-1} I(J)} \int_{-\infty}^{\infty} \frac{d\chi}{2\pi}
\exp[{|\varepsilon|^{-1}\{\varepsilon \overline{v_{\rm adi}^{\chi}} + |\varepsilon| \overline{v_{\rm ex}^{\chi}}+\sum_{j=2}^{\infty} \varepsilon^j \overline{v_{{\rm na},j}^{\chi}} \}
-|\varepsilon|^{-1} g(\chi) }
]
. \label{prob2} 
\end{eqnarray}
Because $g(\chi)$ satisfies $g(\chi_c(J)) = g'(\chi)|_{\chi = \chi_c(J)} = 0$, with the aid of the variable transformation $u= \frac{\chi}{|\chi|} \sqrt{2g(\chi)}$, the saddle point approximation for $P_{\varepsilon}(J)$ for $|\varepsilon|\ll 1$ leads to
\begin{eqnarray}
P_{\varepsilon}(J)  &=& |\varepsilon|^{-1} e^{-|\varepsilon|^{-1}I(J)} \int_{-\infty}^{\infty} \frac{du}{2\pi}\left| \df{\chi}{u} \right| e^{\varepsilon^{-1}\{ \varepsilon \overline{v_{\rm adi}^{\chi}} + |\varepsilon| \overline{v_{\rm ex}^{\chi}}+\sum_{j=2}^{\infty}\varepsilon^j \overline{v_{{\rm na},j}^{\chi}} \}}e^{-|\varepsilon|^{-1} u^2/2 }\nonumber \\
&\simeq&  \frac{1}{\sqrt{2\pi |\varepsilon|^{-1}\partial_{i\chi}^2\overline{\lambda_0^{\chi}}|_{\chi=\chi_c(J)}} } \exp{\left[ -|\varepsilon|^{-1}  \left\{ I(J)-\varepsilon \overline{v_{\rm adi}^{\chi_c(J)}} - |\varepsilon| \overline{v_{\rm ex}^{\chi_c(J)}} +\varepsilon \sum_{j=1}^{\infty} [ \varepsilon^j F_j^{\rm odd}(\chi_c(J)) + |\varepsilon|^j F_j^{\rm even}(\chi_c(J)) ]  \right\}\right]}, \nonumber \\ \label{prob3}
\end{eqnarray}
where we have introduced $F_j^{\rm odd}(\chi_c(J))$ and $F_j^{\rm even}(\chi_c(J))$ as the formal expanding functions of $\varepsilon^j$. 
Here, the explicit expressions of these functions at $j=1$ are given by 
\begin{eqnarray}
F_1^{\rm odd}(\chi_c(J)) &:=& -\overline{v_{na,2}^{\chi_c(J)}}+\left. \frac{1}{\partial_{i\chi}^2\overline{\lambda_0^{\chi}}}\left[\partial_{i\chi}^2\overline{v_{\rm adi}^{\chi}} + (\partial_{i\chi}\overline{v_{\rm adi}^{\chi}})^2 + (\partial_{i\chi}\overline{v_{\rm ex}^{\chi}})^2-\frac{5}{12}\frac{ \partial_{i\chi}^{4}\overline{\lambda_0^{\chi}}}{\partial_{i\chi}^{2}\overline{\lambda_0^{\chi}}} \right]\right|_{\chi=\chi_c(J)}, \label{fodd}\\
F_1^{\rm even}(\chi_c(J)) &:=& \left. \frac{1}{\partial_{i\chi}^2\overline{\lambda_0^{\chi}}}\left[\partial_{i\chi}^2\overline{v_{\rm ex}^{\chi}} + 2(\partial_{i\chi}\overline{v_{\rm adi}^{\chi}}) (\partial_{i\chi}\overline{v_{\rm ex}^{\chi}}) \right]\right|_{\chi=\chi_c(J)}. \label{feven}
\end{eqnarray}
Therefore, the fluctuation theorem is formally written as
\begin{eqnarray}
\lim_{ |\varepsilon| \rightarrow 0 } |\varepsilon |\ln \frac{P_{\varepsilon}(J)}{P_{\varepsilon}(-J)} &=& \varepsilon\{ \overline{v^{\chi_c(J)}}- \overline {v^{\chi_c(-J)} }\} \nonumber \\
& & -\varepsilon \sum_{j=1}^{\infty} \left[ \varepsilon^j \{F_j^{\rm odd}(\chi_c(J)) -F_j^{\rm odd}(\chi_c(-J)) \} + |\varepsilon|^j\{F_j^{\rm even}(\chi_c(J))-F_j^{\rm even}(\chi_c(-J)) \} \right] \nonumber \\
& & +\cdots. \label{formft}
\end{eqnarray}
Unfortunately, this formal expansion is useless, because the explicit expression of $F_j^{\alpha}$ with $\alpha={\rm odd}$ or $\alpha={\rm even}$ for $j\ge 2$ is too complicated.
Instead, let us use the expansion of $P_\varepsilon(J)$ in Eq.\ (\ref{gaussdev}) together with Eq.~\eqref{<J>} we obtain the fluctuation theorem [Eq. (\ref{ft1})].

\subsection{Nonequilibrium relations: FDR and the reciprocal relations}

In this section, we derive relations among cumulants. 
We formally expand cumulants in terms of $\varepsilon$ against the symmetries (\ref{adpar}) and (\ref{napar}) :
\begin{eqnarray}
\expec{J^n}_c &:=& 
\frac{(2\pi)^{n-1}}{(\Gamma \omega_0 \tau_p)^n} \left. \frac{\partial^n}{\partial (i\chi)^n} S_{\tau_p}(\chi) \right|_{\chi =0} 
=  |\varepsilon|^{n-1} \sum_{j = 1}^{\infty} G_{n, 2j - 1} \varepsilon^{2j - 1} \ \ \ ({\rm for}\ n : {\rm odd}), \label{ir1}\\
\expec{J^n}_c &:=&
 \frac{(2\pi)^{n-1}}{(\Gamma \omega_0 \tau_p)^n} \left. \frac{\partial^n}{\partial (i\chi)^n} S_{\tau_p}(\chi) \right|_{\chi =0} 
 = |\varepsilon|^{n-1} \left[ G_{n,0} +G_{n,1}|\varepsilon| + \sum_{j = 1}^{\infty} G_{n, 2j } \varepsilon^{2j } \right] \ \ \ ({\rm for}\ n : {\rm even}),\label{ir2}
\end{eqnarray}
where we have introduced the $n$th cumulant $\expec{J^n}_c$ 
and used the symmetry relation $J \rightarrow -J$ under the transformation $\Omega\to -\Omega$.
Integrating fluctuation theorem (\ref{ft1}) and the balance of terms at $O(|\varepsilon|)$ leads to the FDR as
\begin{eqnarray}
G_{20} = \frac{2}{A_1}G_{11} \label{fdr1}
\end{eqnarray}
To illustrate the equivalency between Eqs. (\ref{fdr1}) and the standard FDR [Eq. (\ref{sfdr1})], we rewrite Eq.\ (\ref{fdr1}) as 
\begin{eqnarray}
G_{20} = 2\tilde G_{11}, \label{fdr12}
\end{eqnarray} 
where we have introduced $\tilde{G}_{11}:=G_{1n}/A_1$.
Note that this FDR is a relation in the adiabatic process, because the relation is derived from that in the lowest order of $|\varepsilon|$.
 In other words, the higher order relations among cumulants are the results of nonadiabatic effects. 

Now, let us discuss the relation between cumulants at $O(\varepsilon^2)$. 
The relation for $G_{21}$ [derived in Eq.\ (\ref{apft22})] is given as
\begin{eqnarray}
G_{21} = -\frac{2}{A_1^3} ( A_1BG_{11} + 6A_3G_{11}^2). 
\label{fdr2}
\end{eqnarray} 
Relation (\ref{fdr2}) is completely different from the conventional reciprocal relation Eq.\ (\ref{sfdr2}) for the steady current.
Note that $G_{12}$ is always zero because of the symmetry between $J$ and $-J$, but $G_{21}$ should be finite to be consistent with Fig. 1 (b). 
It is also remarkable that the extended reciprocal relation, Eq.\ (\ref{fdr2}), is the one among nonadiabatic coefficients.

\subsection{Application to a spin-boson system}

In this section, we numerically verify the theoretical results obtained in the previous subsections. 
This enables us to discuss the FDR and the reciprocal relation in the spin-boson system.


From Fig. 1(a) we evaluate $G_{11}/2\pi =4.77 \times 10^{-5}$.
From Fig. 1(b) $G_{20}$ is estimated as $G_{20}/2\pi = 2.38 \times 10^{-3}$ and $G_{21}/2\pi = 1.77 \times 10^{-4}$.
Then, we verify the validity of Eq.\ (\ref{fdr1}) because of the relation $2G_{11}/G_{20} \simeq 4.00 \times 10^{-2}\simeq A_1$. 

To verify Eq.~\eqref{fdr2} let us rewrite
$B$ as 
\begin{eqnarray}
B = -4 \left. \frac {(\partial_{i\chi}\overline{v_{\rm adi}^{\chi}}) (\partial_{i\chi}^2\overline{v_{\rm ex}^{\chi}}) }{(\partial_{i\chi}^2\overline{\lambda_0^{\chi}})^2}\right|_{\chi = 0} = -4 \frac{G_{11}G_{21}}{G_{20}^2} \label{brep}
\end{eqnarray}
with the aid of Eqs.\ (\ref{fodd}) and (\ref{feven}). 
Then, $B$ is evaluated as $B = -9.49 \times 10^{-6}$.
The non-Gaussian parameter $b(0)$ introduced in Eq. \eqref{gaussdev} is estimated as $b(0)\simeq 5.4\times 10^6$ as written in the caption of Fig. \ref{fig4}. 
Then, 
$A_3$ introduced in Eq.\ (\ref{cob}) becomes $A_3 = 2.08 \times 10^{-6}$.
As a result, the right hand side of Eq.\ (\ref{fdr2}) is $8.96 \times 10^{-10}$, which is nearly two times larger than the value of the left hand side, $G_{21} = 4.48 \times 10^{-10}$.
Although we have not verified the quantitative validity of Eq.~\eqref{fdr2} from our analysis, the violation of the reciprocal relation is clear because of $G_{12}=0$ and $G_{21}\ne 0$.

\section{Discussion and Conclusion}

Let us discuss our results.
(i) From our simulation of the spin-boson system, we could not quantitatively verify Eq.\ (\ref{fdr2}) because of the numerical difficulty for the calculation in the nonadiabatic region.
Moreover, it is difficult to evaluate the higher order cumulants precisely. 
Such difficulty has been reported even in an actual experiment for the steady fluctuation theorem~\cite{Nakamura}.
We also try to estimate the expansion coefficients theoretically in terms of the asymptotic expansion as developed in Ref. \cite{watanabe}.
(ii) Our theory includes only one kind of current. 
If there are multiple currents, Onsager's reciprocal relation becomes a relation between linearly nonequilibrium currents, which is quite different from the reciprocal relation for a nonadiabatic relation in our system under a single current.
Therefore we will have to study such systems in the near future. 
(iii) In this paper, we have considered the sufficiently slow modulation of parameters. 
On the other hand, Floquet theory\cite{Tannor} can be applied if the operation speed is sufficiently fast. 
There are few studies under intermediate operation speed in which non-Markovian effects should be relevant.
(iv) Although our framework does not assume any specific form of a closed path of parameters, our numerical simulation is only performed under sinusoidal operations.
If we discuss the thermodynamics of this system, we need to analyze a system including fast operations which cause the change of the reservoirs. 
Such a problem will have to be analyzed in the near future.
Note that path dependent entropy exists in our setup\cite{Sagawa, Yuge2013} and thus the corresponding thermodynamics should be nontrivial. 
(v) We have not assumed that the target system is small in our framework, but we can only treat small systems for the actual numerical analysis. If the many-body effects in the system become relevant, we may find some qualitatively new results that are not discussed in this paper. 
(vi) The original geometric pump was proposed by Thouless\cite{Thouless} in a closed system. Therefore, we will have to analyze the geometric fluctuation theorem even for a closed system.
(vii) We have introduced a singular term in the expansion of the cumulant-generating function in Eq.\ (\ref{geoexp}). 
This singular term reflects on the singularity of $\langle J^2 \rangle_c$ but we have not identified its origin. We will have to clarify the mechanism of this singularity in the near future.
 
In conclusion, we have a fluctuation theorem, the geometric fluctuation theorem for a spin-boson system under the influence of multiple periodically modulated  parameters. If we can ignore the contribution of the dynamical phase, we derive the FDR from the geometric fluctuation theorem, while the conventional reciprocal relation is no longer valid but the 
alternative relation exists.
The validity of the FDR and the violation of the reciprocal relation has been verified through the Monte Carlo simulation, though we could not get quantitative agreement between the theory and the simulation for the alternative relation.
 
\begin{acknowledgements} 
We thank Yu Watanabe, Ryosuke Yoshii, and Keiji Saito for various discussions and their useful comments. 
One of the authors thanks Satoshi Takada for his technical help on the occasion of the revision of this manuscript.
The numerical calculations were carried
out on a CRAY XC40 at YITP in Kyoto University. 
Numerical computation in this work was carried out at the Yukawa Institute  
Computer Facility.
This work is partially supported by Grant-in-Aid of MEXT for Scientific Research (Grant No. 16H04025).
\end{acknowledgements}

\appendix

\section{The rate function and the saddle point method}
In this appendix, we introduce the rate function under a time modulation and explain the saddle point method to evaluate the rate function. We also briefly summarize their properties.

First, let us introduce the the time dependent rate function for $\lambda_0^{\chi}(t)$ : 
\begin{eqnarray}
I(J,t) := \max_{\chi}[i\chi J -\lambda_0^{\chi}(t)] = i\chi_c(J,t) J - \lambda_0^{\chi_c(J,t)}(t), \label{rf1}
\end{eqnarray}
where $\chi_c(J,t)$ is obtained by solving $J= \partial_{i\chi} \lambda_0^{\chi}(t)|_{\chi=\chi_c}$ because of the convexity of $\lambda_0^{\chi}(t)$.
According to the symmetry relation [Eq.\ (\ref{sym})], $\chi_c(J,t)$ and $I(J,t)$, respectively, satisfy
 \begin{eqnarray}
\chi_c(-J,t) = -\chi_c(J,t) + i\Delta \alpha(t), \label{chisym}
\end{eqnarray}
and
\begin{eqnarray}
I(-J,t) = I(J,t) + \Delta \alpha(t) J. \label{Isym}
\end{eqnarray}

Next, we introduce the time-averaged rate function for $\overline{\lambda_0^{\chi}}$ :
\begin{eqnarray}
I(J) := \max_{\chi}[i\chi J -\overline{\lambda_0^{\chi}}] =i\chi_c(J) J - \overline{\lambda_0^{\chi_c(J)}}, \label{rf2}
\end{eqnarray}
where $\chi_c(J)$ is obtained by solving $J=\partial_{i\chi} \overline{\lambda_0^{\chi}}|_{\chi=\chi_c}$.
 
From Eq.\ (\ref{rf2}) and the even property of $\lambda^{\chi}(t)$ against $\chi-i\Delta \alpha(t)/2$, we obtain
\begin{eqnarray}
\chi_c(J) &=& \overline{\chi_c(J,t)}, \\
\overline{\lambda_0^{\chi_c(J)}} &=&  \overline{  \lambda_0^{\chi_c(J,t)}(t)}.
\end{eqnarray}
When we adopt the condition $\overline{\Delta \alpha(t)} =0$, we obtain the relations
\begin{eqnarray}
\chi_c(J) &=& -\chi_c(-J), \label{chisym}\\
\overline{\lambda_0^{\chi_c(J)}} &=&\overline{  \lambda_0^{\chi_c(-J)}}, \label{lamev}\\
I(J) &=& I(-J). \label{ratesym}
\end{eqnarray}

\section{Formulas in The Spin-Boson System}
In this appendix, we apply our general argument to a spin-boson system. 

The eigenvalue with the maximum real part and the corresponding left and right eigenvectors of $\mathcal K^{\chi}_{\vec{\alpha}(t)}$ are explicitly written as
\begin{eqnarray}
\lambda^{\chi}_0 (t) &=& \frac{a_1(t)+a_4(t)}{2} + \sqrt{\left(\frac{a_1(t)-a_4(t)}{2}\right)^2 + a^{\chi}_2(t) a^{\chi}_3(t) }, \label{lambda} \\
\label{39}
\kket{r_0^{\chi}(t) } &=& \frac{1}{N^{\chi}(t)} \begin{pmatrix}
1 \\
\displaystyle \frac{\lambda^{\chi}_0 (t)-a_1(t)}{a_2^{\chi}(t)}
\end{pmatrix} , \\
\bbra{l^{\chi}_0(t)} &=& \begin{pmatrix}
1 & \displaystyle \frac{\lambda^{\chi}_0 (t)-a_1(t)}{a_3^{\chi}(t)}
\end{pmatrix}, \label{lefte}
\end{eqnarray}
where we have introduced
\begin{eqnarray}
N^{\chi}(t) = 1+ \frac{(\lambda^{\chi}_0 (t)-a_1(t))^2}{a_2^{\chi}(t)a_3^{\chi}(t)}.
\end{eqnarray}
This $\lambda_0(t)$ has the symmetry $(\ref{sym})$ with $\Delta \alpha(t) = \ln{[n_L(1+n_R)/\{ n_R(1+n_L)\}]}
= \hbar\omega_0(\beta_L(t) -\beta_R(t))$ because $a_2^{\chi}(t)a_3^{\chi}(t)$ has the following symmetry relation :
\begin{eqnarray}
a_2^{\chi}(t)a_3^{\chi}(t) &=& \Gamma^2n_L(t)[1+n_L(t)] + \Gamma^2n_R(t)[1+n_R(t)]\nonumber \\
& & + \Gamma^2 n_L(t)[1+n_R(t)]e^{i\chi\hbar\omega_0} + \Gamma^2n_R(t)[1+n_L(t)]e^{-i\chi\hbar\omega_0} \nonumber \\
&=& a_2^{-\chi+i\Delta \alpha(t) }(t)a_3^{-\chi+i\Delta \alpha(t) }(t) .
\end{eqnarray}

\section{The derivations of the FDR and the reciprocal relation }
In this appendix, we briefly exemplify the FDR and reciprocal relation under a steady current in terms of the steady fluctuation theorem in the first part.
In the second part, we also illustrate the existence of the FDR and the absence of the reciprocal relation for a geometric pumping process.

\subsection{Steady fluctuation theorem and its relations among cumulants }
In this section, we review the conventional FDR and the reciprocal relation from the steady fluctuation theorem\cite{SaitoU}.
The steady fluctuation theorem is the direct consequence of the LLGC symmetry: 
\begin{eqnarray}
\lambda_0^{\chi} = \lambda_0^{-\chi+i\mathcal A}, \label{sftlam}
\end{eqnarray}
where we have introduced the affinity $\mathcal A$ such that $\mathcal A = \beta \Delta \mu$ if the input is a constant voltage $\Delta \mu$ and the temperature of environments is $\beta^{-1}$. 
The rate function $I(J) := \max_{\chi}\{ i\chi - \lambda_0^{\chi} \}$ has the symmetry 
\begin{eqnarray}
I(J) - I(-J) = -\mathcal A J. \label{sisym}
\end{eqnarray}
By using $I(J)$, the steady probability distribution is written as
\begin{eqnarray}
\lim_{\tau \rightarrow \infty} P_{\tau}(J) \sim e^{-\tau I(J)}.   
\end{eqnarray}
Therefore, the fluctuation theorem with the distribution form is given by
\begin{eqnarray}
\lim_{\tau \rightarrow \infty} \frac{1}{\tau} \ln \frac{P_{\tau}(J)}{P_{\tau}(-J)}= \mathcal A J, \label{sft}
\end{eqnarray}

We expand cumulants for $\mathcal A$ :
\begin{eqnarray}
\tau^{n-1} \expec{J^n}_c = \sum_{m} L_{n m}\frac{\mathcal A^m}{m!}.
\end{eqnarray}
Note that this expansion is valid if the affinity is small.
Let the integral fluctuation theorem
$\int_{-\infty}^\infty dJ P(J)e^{-\tau \mathcal{A} J}=1$
 also be expanded for $\mathcal A$ :
\begin{eqnarray}
1 &=& \int_{-\infty}^{\infty}dJ P(J) e^{-\tau \mathcal A J} \nonumber \\
&=& 1 +\sum_{n=1} (-1)^n \frac{\mathcal A^n\tau^n}{n!}\expec{J^n} \nonumber \\
&=& 1 - \mathcal A \tau \expec{J} + \frac{\mathcal A^2 \tau ^2}{2} [\expec{J^2}_c + \expec{J}^2] - \frac{\mathcal A^3 \tau ^3}{3!} [\expec{J^3}_c + 3\expec{J^2}_c\expec{J} + \expec{J}^3] +  \cdots 
\nonumber \\
& & + \frac{\mathcal A^4 \tau ^4}{4!} [\expec{J^4}_c + 4\expec{J^3}_c\expec{J} +3\expec{J^2}_c^2+6\expec{J^2}_c\expec{J}^2+ \expec{J}^4] \nonumber \\
&=& 1 - \mathcal A \tau \left[L_{11} \mathcal A + L_{12} \frac{\mathcal A^2}{2} + L_{13}\frac{\mathcal A^3}{3!} +\cdots \right] + \frac{\mathcal A^2 \tau }{2} [L_{20} + L_{21}\mathcal A +\cdots +\tau (L_{11}\mathcal A+\cdots)^2] \nonumber \\
& & -\frac{\mathcal A^3 \tau }{3!} [L_{31}\mathcal A +\cdots +3 \tau (L_{20} +\cdots )(L_{11} \mathcal A + \cdots) + \tau^2(L_{11} \mathcal A +\cdots )^3] \nonumber \\
& & + \frac{\mathcal A^4 \tau}{4!} [ ( L_{40} + \cdots ) + 4\tau (L_{31} \mathcal A +\cdots)( L_{11} \mathcal A +\cdots) + 3\tau^2 (L_{20} + \cdots)^2 \nonumber \\
& & + 6\tau^2(L_{20} + \cdots)(L_{11}\mathcal A + \cdots)^2 +\tau^3 (L_{11}\mathcal A + \cdots)^4 ] \nonumber \\
\end{eqnarray}
Then, we obtain the FDR
\begin{eqnarray}
L_{20} = 2 L_{11}
 \label{sfdr1}
\end{eqnarray}
at $O(\mathcal A^2\tau)$, 
and the reciprocal relation
\begin{eqnarray}
L_{21} = L_{12}
 \label{sfdr2}
\end{eqnarray}
at $O(\mathcal A^3\tau)$.
We can also derive nonlinear relations as higher order corrections.

\subsection{Derivation of relations among cumulants for geometric pumping}
In this section, we derive Eqs.\ (\ref{fdr1}) and (\ref{fdr2}) for the geometric pumping process. Rewriting Eq.\ (\ref{ft1}) as $P_{\varepsilon}(-J)=P_{\varepsilon}(J)e^{-[A_1 J + A_3\tilde  J^3 +\cdots + |\varepsilon| B J +\cdots ] \varepsilon/|\varepsilon|}$, we obtain the expansion series in terms of $\varepsilon$ as
\begin{eqnarray}
1 &=& \int_{-\infty}^{\infty}dJ P_{\varepsilon}(J)e^{-[A_1 J + A_3\tilde  J^3 +\cdots  + |\varepsilon| B J + \cdots] \varepsilon/ |\varepsilon|} \nonumber \\
&=& 1 - \frac{\varepsilon}{ |\varepsilon|} (A_1+ |\varepsilon| B ) \expec{J} + \frac{(A_1+ |\varepsilon| B )^2 }{2}\expec{J^2} - \frac{\varepsilon}{ |\varepsilon|} \left[\frac{(A_1+ |\varepsilon| B )^3}{3!}  + A_3 \right] \expec{J^3} \nonumber \\
& & +  \left[  \frac{(A_1+ |\varepsilon| B )^4}{4!} + (A_1+ |\varepsilon| B ) A_3  \right]\expec{J^4}+\cdots \label{intcal}
\end{eqnarray}
Here, each moment is rewritten as cumulants
\begin{eqnarray}
\expec{J} &=& G_{11} \varepsilon +  \cdots \\
\expec{J^2} &=& \expec{J^2}_c + \expec{J}^2 =  |\varepsilon|( G_{20}+ G_{21} |\varepsilon| + G_{22} \frac{\varepsilon^2}{2} +\cdots ) + (G_{11}\varepsilon + \cdots ) ^2\nonumber \\
& =& G_{20} |\varepsilon|  + (G_{21} + G_{11}^2) \varepsilon^2 + \cdots, \\
\expec{J^3} &=& \expec{J^3}_c+3\expec{J^2}_c\expec{J}+\expec{J}^3 =  |\varepsilon|^2 (G_{31}\varepsilon + \cdots ) + 3 |\varepsilon| (G_{20} + G_{21} |\varepsilon| + \cdots ) (G_{11} \varepsilon + \cdots) + (G_{11} \varepsilon + \cdots)^3 \nonumber \\
&= & 3G_{11}G_{20}  |\varepsilon|\varepsilon + \cdots, \\
\expec{J^4} &=& \expec{J^4}_c + 4\expec{J^3}_c\expec{J} +3\expec{J^2}_c^2+6\expec{J^2}_c\expec{J}^2+ \expec{J}^4\nonumber \\
&=&  |\varepsilon|^3(G_{40} + \cdots) + 4 |\varepsilon|^2(G_{31}\varepsilon+\cdots)(G_{11}\varepsilon + \cdots) + 3  |\varepsilon|^2(G_{20} +G_{21} |\varepsilon|+ \cdots)^2\nonumber \\
& & + 6 |\varepsilon|(G_{20} +G_{21} |\varepsilon| + \cdots)(G_{11} \varepsilon + \cdots)^2 + (G_{11}\varepsilon + \cdots)^4\nonumber \\
&=& 3G_{20}^2  \varepsilon^2 + \cdots, 
\end{eqnarray}
where we have used Eqs.\ (\ref{ir1}) and (\ref{ir2}) to obtain the final expression. 
Then, Eq.\ (\ref{intcal}) is rewritten as 
\begin{eqnarray}
0 &=& \varepsilon \left[- A_1G_{11} + \frac{A_1^2}{2}G_{20} \right] \nonumber \\
& &+ \varepsilon^2 \left[ -BG_{11} + \frac{A_1^2}{2} (G_{21} + G_{11}^2) + A_1BG_{20} - 3\left( \frac{A_1^3}{3!} + A_3\right)G_{11}G_{20} + 3\left(\frac{A_1^4}{4!} + A_1A_3 \right)G_{20}^2\right]+ \cdots. \label{oexp}
\end{eqnarray}
From the first order for $\varepsilon$ in Eq.\ (\ref{oexp}), we obtain
\begin{eqnarray}
G_{20} = \frac{2}{A_1} G_{11}. \label{apft1}
\end{eqnarray}
From the second order for $\varepsilon$ in Eq.\ (\ref{oexp}), we obtain
\begin{eqnarray}
0 = -BG_{11} + \frac{A_1^2}{2} (G_{21} + G_{11}^2) + A_1BG_{20} - 3\left( \frac{A_1^3}{3!} + A_3\right)G_{11}G_{20} + 3\left(\frac{A_1^4}{4!} + A_1A_3 \right)G_{20}^2. \label{apft2}
\end{eqnarray}
With the aid of the FDR between $G_{20}$ with $G_{11}$ in Eq.\ (\ref{apft1}), Eq.\ (\ref{apft2}) is rewritten as
\begin{eqnarray}
G_{21} = -\frac{2}{A_1^3} ( A_1BG_{11} + 6A_3G_{11}^2). \label{apft22}
\end{eqnarray}

\end{document}